\documentclass{article}
\usepackage[utf8]{inputenc}
\usepackage{lscape,epsfig,graphicx,subcaption}
\usepackage[section]{placeins}
\usepackage{makecell}
\usepackage{multirow}
\usepackage{siunitx,amssymb}
\usepackage{authblk}
\usepackage{color}
\usepackage{url}
\usepackage{overpic}
\usepackage {lineno}
\usepackage{ulem}

\title{Mass production and performance study on the 20-inch PMT acrylic protection covers in JUNO}
\author[a]{Miao He\thanks{hem@ihep.ac.cn}}
\author[a]{Zhonghua Qin}
\author[a]{Diru Wu}
\author[a]{Meihang Xu}
\author[a]{Wan Xie}
\author[a]{Fang Chen}
\author[a]{Xiaoping Jing}
\author[b]{Genhua Yin}
\author[b]{Shengjiong Yin}
\author[b]{Linhua Gu}
\author[b]{Xiaofeng Xia}
\author[b]{Qinchang Wang}

\affil[a]{Institute of High Energy Physics, Beijing 100049, China}
\affil[b]{Zhejiang Huashuaite New Material Technology Co.,Ltd, Jiaxing, 314311, China}

\begin{document}

\maketitle

\begin{abstract}
The Jiangmen Underground Neutrino Observatory is a neutrino experiment that incorporates 20,012 20-inch photomultiplier tubes (PMTs) and 25,600 3-inch PMTs. A dedicated system was designed to protect the PMTs from an implosion chain reaction underwater. As a crucial element of the protection system, over 20,000 acrylic covers were manufactured through injection molding, ensuring high dimensional precision, mechanical strength, and transparency. This paper presents the manufacturing technology, mass production process, and performance characteristics of the acrylic covers.
\end{abstract}

\section{Introduction}

The photomultiplier tube (PMT) is a photon detector consisting of an evacuated glass shell coated with a photocathode and equipped with an electron multiplier. In large-volume neutrino experiments, PMTs are usually situated in water such as in the case of Super-Kamiokande~\cite{Super-Kamiokande:2002weg}, or organic liquids, such as seen in Borexino~\cite{Borexino:2008gab} and KamLAND~\cite{KamLAND:2002uet}. Prolonged hydrostatic pressure can lead to the breakage of the glass shell, causing a shockwave when water rushes into the vacuum and then reverse-ejects. If the energy of the shockwave is sufficient to break neighboring PMTs, a secondary shockwave is generated, leading to a chain reaction. In 2001, over 7,000 PMTs were imploded due to a cascade shockwave in Super-Kamiokande. To protect PMTs from cascade implosion, one practical approach is to encapsulate the PMT in a stronger shell with small inlets. When one PMT is broken, the protective shell slows down the water, reducing the energy of the shockwave far below the implosion threshold of neighboring PMTs. This design was implemented in the fully rebuilt Super-Kamiokande in 2006~\cite{Super-Kamiokande:2010tar}.

\begin{figure}[ht]
\centering
  \includegraphics[height=0.3\textwidth]{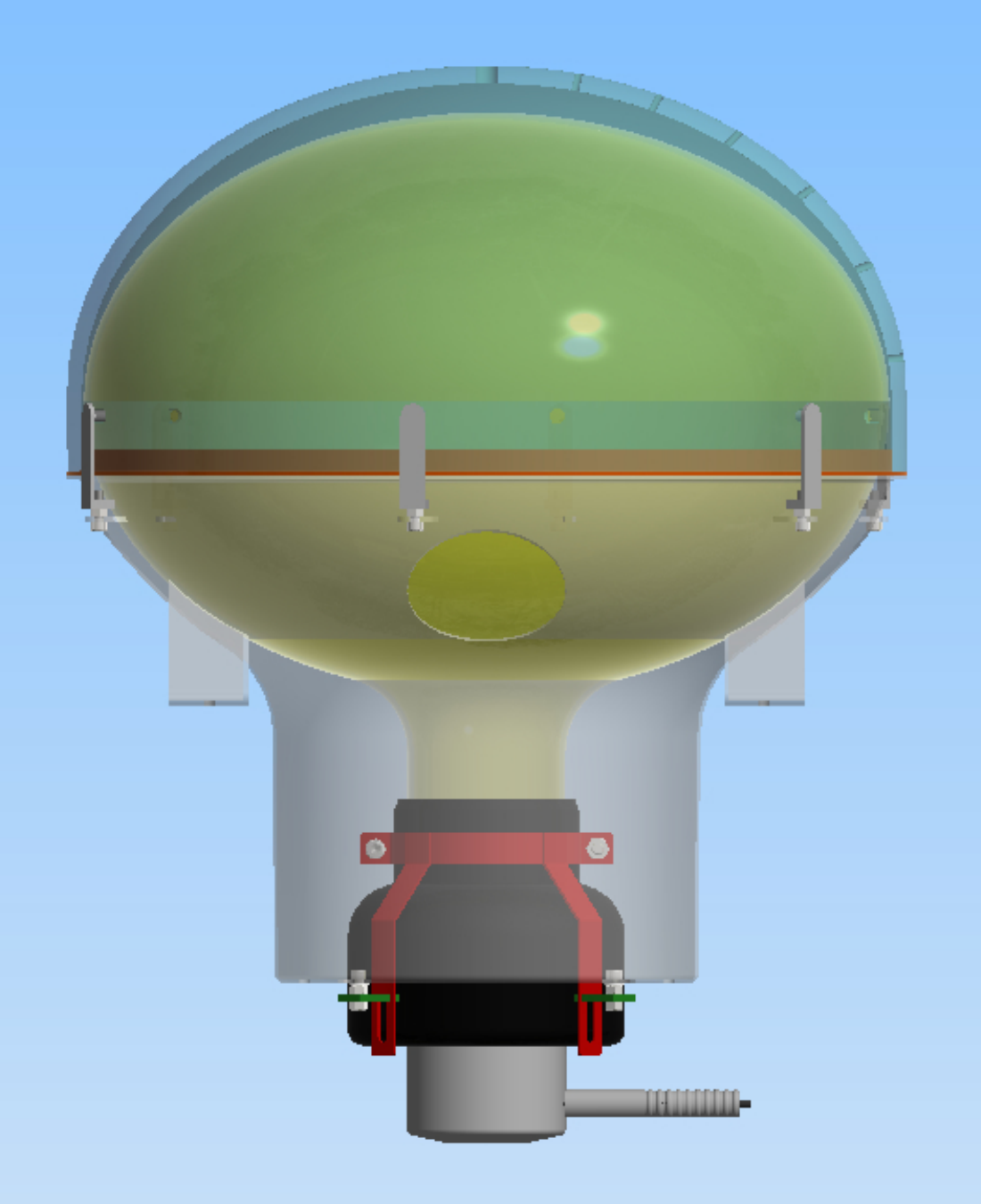}
  \includegraphics[height=0.3\textwidth]{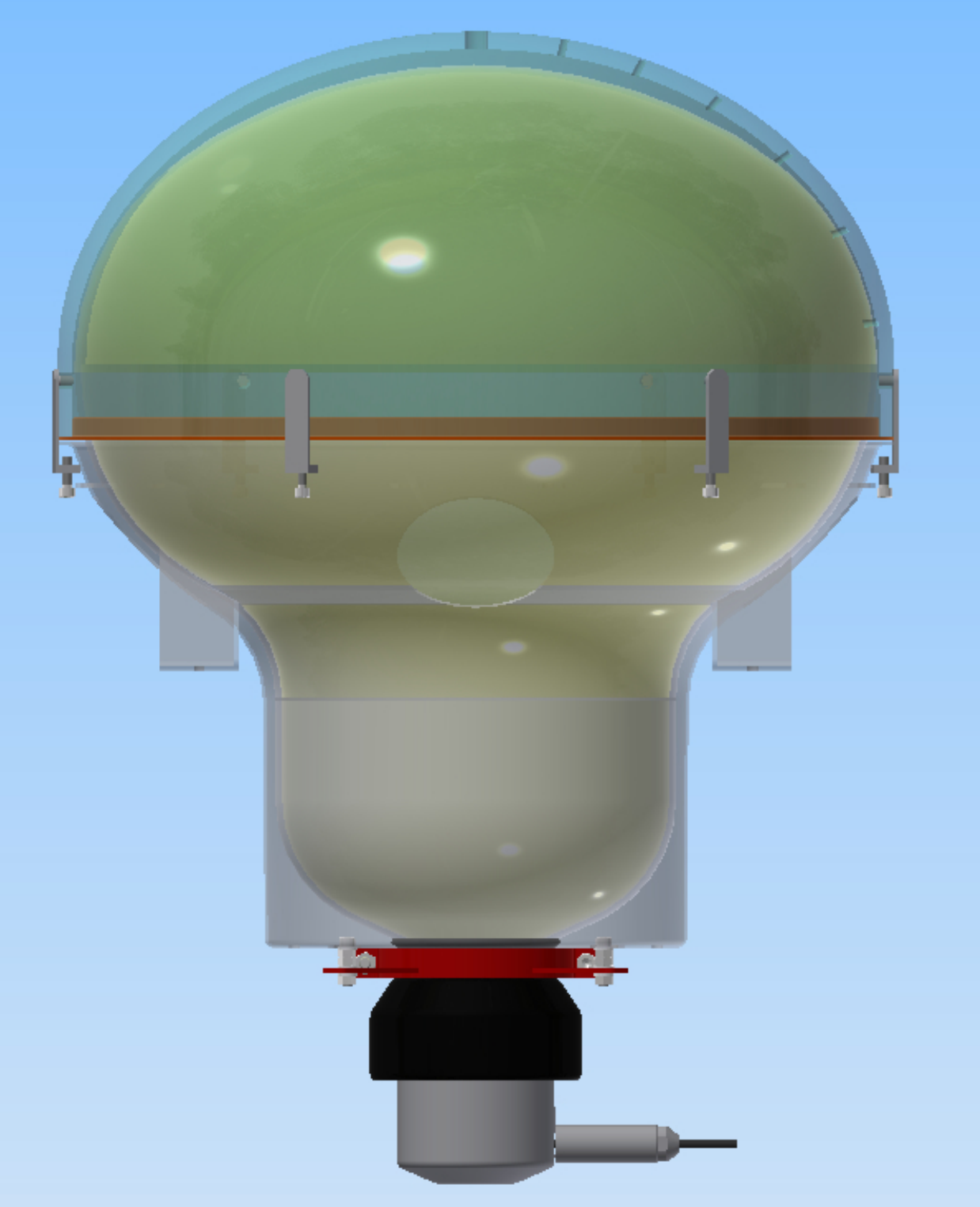}
  \includegraphics[height=0.3\textwidth]{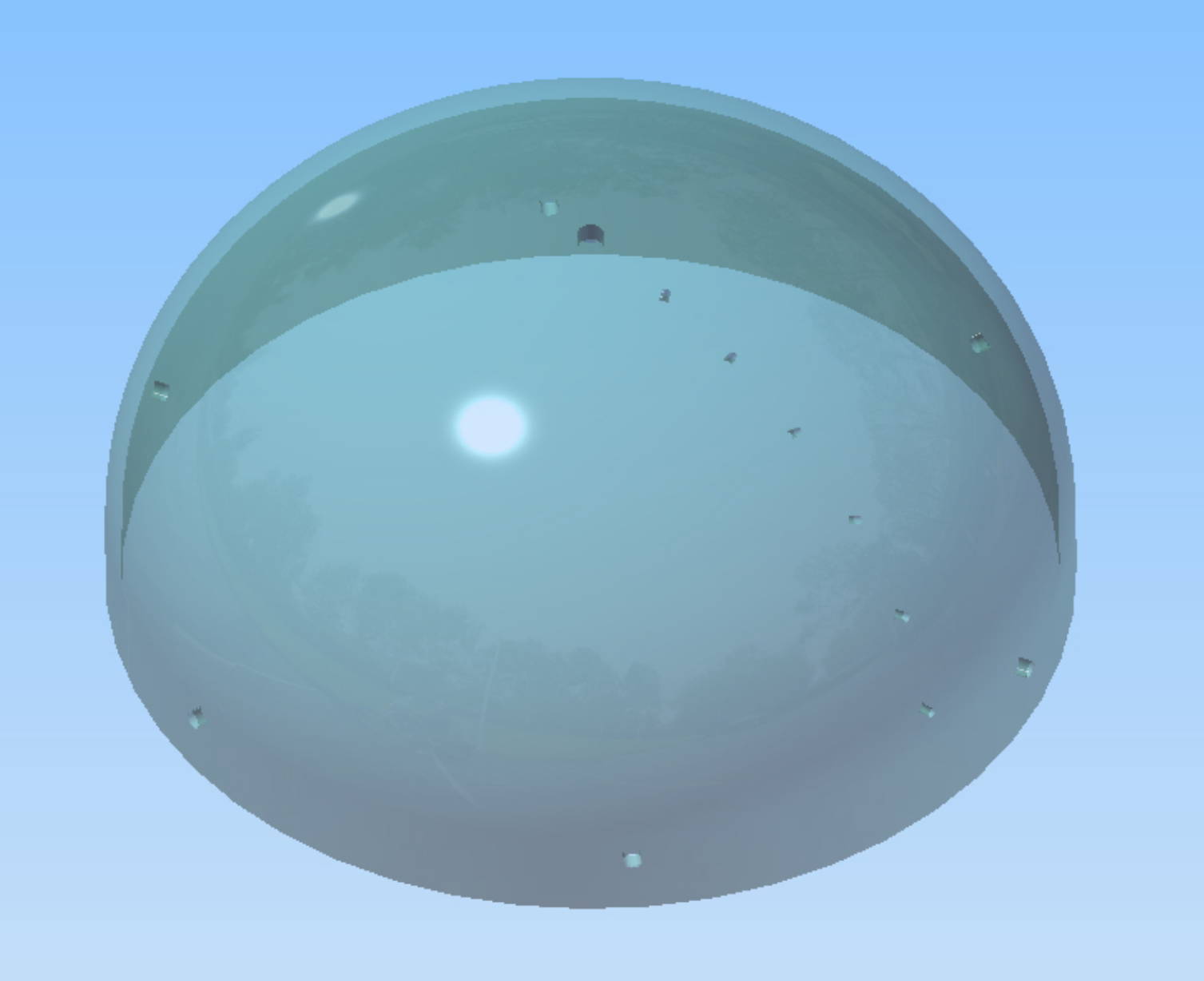}
\caption{Design of the protection system. The left figure is the NNVT PMT and the middel figure is the HPK PMT, both contained in the same protection system, consisting of an acrylic top cover, a stainless steel bottom cover and six connecting hooks. The right figure is the acrylic cover equipped on top of the PMT. The 15-mm diameter hole on the top and the rest six 5-mm diameter holes along a meridian are used as air vents. The six 10-mm diameter holes 30-mm far from the bottom are used for the connecting hooks.}
\label{fig:design}
\end{figure}

The Jiangmen Underground Neutrino Observatory (JUNO)~\cite{An:2015jdp, Djurcic:2015vqa, JUNO:2022hxd} is a neutrino experiment currently being constructed in southern China. The JUNO detector contains 20,012 PMTs with a diameter of 508~mm (20~inches), all of which are submerged in water up to a depth of 45~m. Among these, 17,612 are closely packed with a minimum clearance of 25~mm and face inward, detecting light from the 20 kton liquid scintillator in the Central Detector (CD), and are therefore referred to as CD PMTs. The remaining 2,400 PMTs face outward, detecting light in the Water Cherenkov detector for the cosmic muon veto, and are thus referred to as veto PMTs. The distance between two veto PMTs, or between CD PMTs and veto PMTs, is greater than 1~m. Additionally, 25,600 PMTs with a diameter of 80~mm (3.1~inches) are installed in the gap between CD PMTs with a minimum clearance of 13.7~mm. Out of the 20-inch PMTs, 15,012 are supplied by North Night Vision Technology Co., Ltd (NNVT) and the remaining 5,000 by Hamamatsu Photonics K.K. (HPK). As shown in Fig.~\ref{fig:design}, the PMT glass shell is a truncated ellipsoid, with the top half being transparent to receive light and the bottom half connected with a neck containing electron multiplier and signal readout components. A waterproofing structure protects the end of the neck.

The design of the protection system for JUNO PMTs was detailed in a previous report~\cite{He:2022qzj} and is illustrated in Fig.~\ref{fig:design}. Each 20-inch PMT will be outfitted with a pair of protection covers capable of accommodating both NNVT PMTs and HPK PMTs. The top half of the cover is made of transparent acrylic, while the bottom half is composed of stainless steel. These two parts are connected by six hooks through six 10~mm diameter holes close to the bottom of the acrylic cover. Additionally, there are seven holes along one of the meridians with a diameter of either 5~mm or 15~mm, serving as air vents during water filling. The inner diameter of the acrylic cover is 512~mm, and the inner height is 200~mm. To withstand a step load of 50~m water pressure and fit the 25~mm clearance, a nonuniform thickness was designed and achieved through injection molding, with 11~mm on the top and 9~mm on the bottom. Consequently, there is a 2~mm gap between the PMT and the acrylic cover, and a 3~mm gap between two adjacent acrylic covers. Therefore, precise control of the thickness and dimensions is critical for both the mechanical performance and the assembly of the acrylic cover. Additionally, the transparency of the acrylic cover is also important as it directly impacts light detection.

This paper presents the manufacturing technology, mass production, and performance study of over 20,000 acrylic covers. The stainless steel covers, currently in the midst of mass production, will not be discussed in this paper. We discuss the major technical challenges, manufacturing procedures, and pilot production in Section~\ref{sec.pilot}. Section~\ref{sec.massprod} covers mass production, mechanical and optical performance studies, and validation through underwater experiments. The conclusion will be provided in Section~\ref{sec.summary}.

\section{Manufacture technology and pilot production}
\label{sec.pilot}

\subsection{Injection molding}

Injection molding is a commonly used forming process for synthetic resins, including acrylic, also known as Polymethylmethacrylate (PMMA). This process involves melting casting acrylic pellets at a temperature exceeding 200~$^{\circ}$C and injecting them into a steel mold at a pressure of approximately 100~MPa. After a few minutes of solidification, the product is removed from the mold.

\begin{figure}[ht]
\centering
  \includegraphics[width=0.5\textwidth]{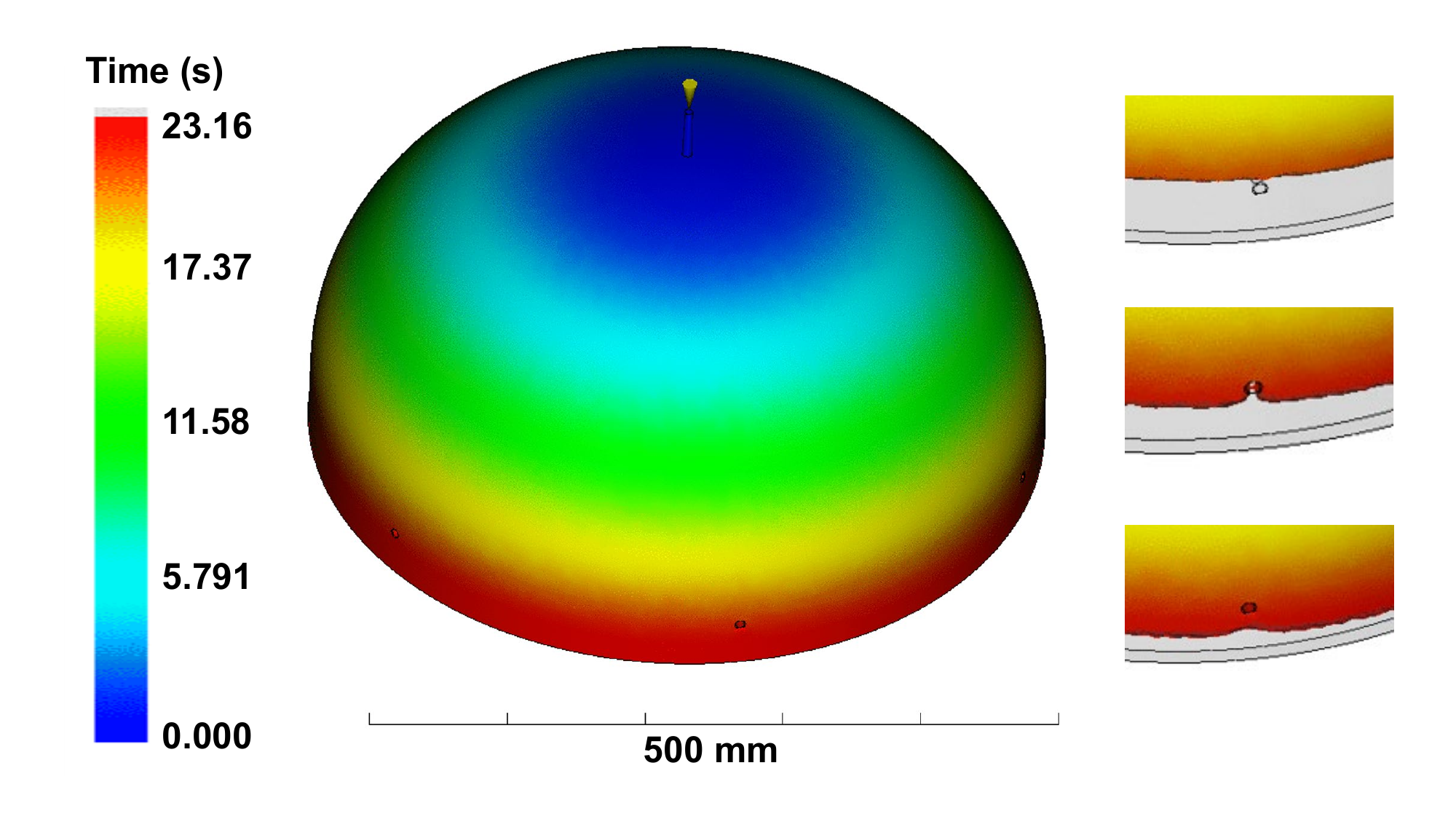}
\caption{Filling time of the acrylic cover obtained from the injection molding analysis. The right panels are zoom-in of one connection hole to demonstrate the weld line.}
\label{fig:injection_ana}
\end{figure}

\begin{figure}[ht]
\centering
  \includegraphics[width=0.5\textwidth]{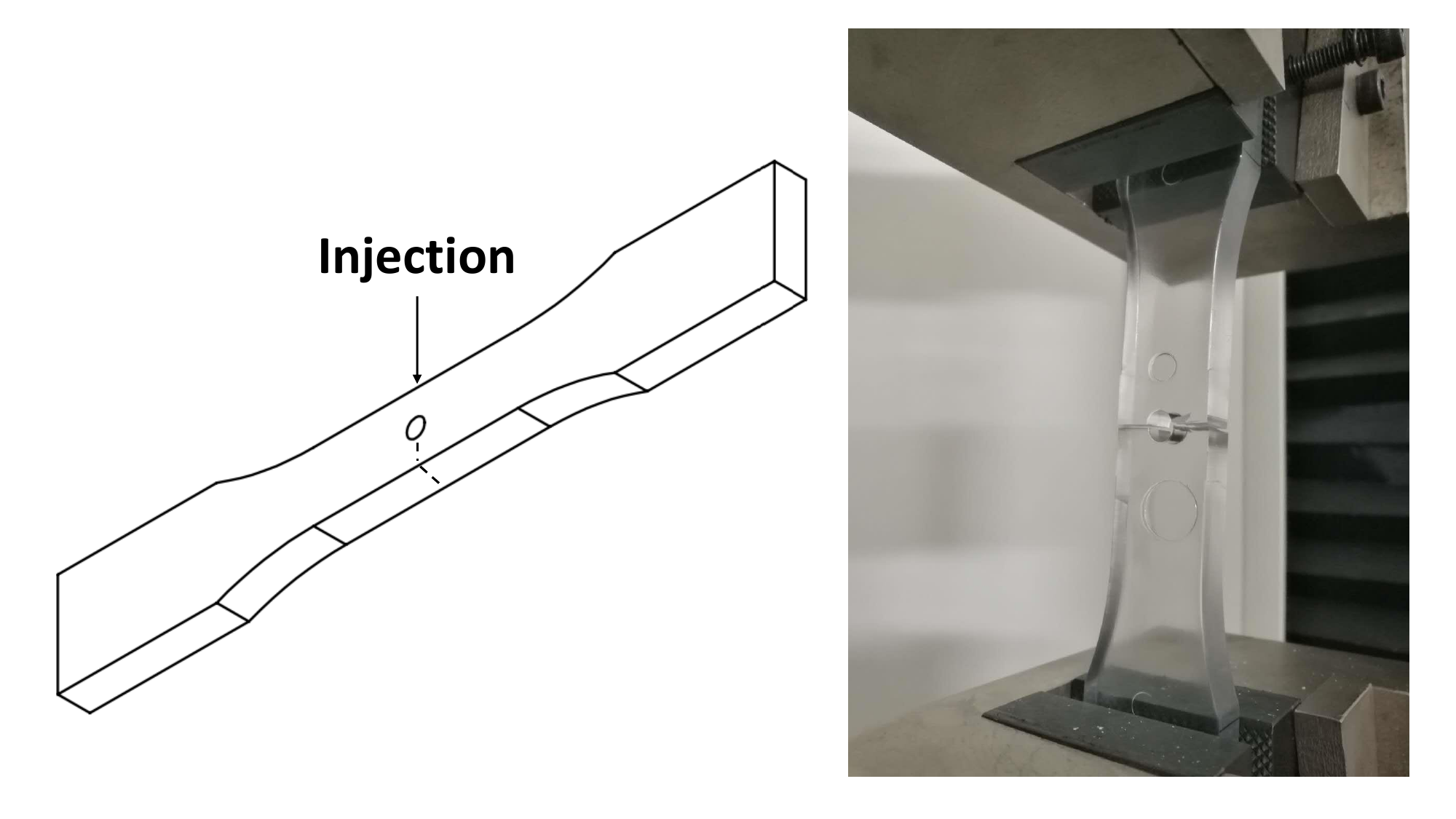}
\caption{Mechanical performance experiment of the weld line. The left panel shows the design of the strip with the arrow indicating the entrance and direction of injection and the dash line presenting the weld line. The right panel shows the tensile test of one strip. }
\label{fig:weld_line}
\end{figure}

A comprehensive mold flow analysis was conducted using typical injection molding parameters to aid in mold design. Due to the symmetric geometry, the injection gate was positioned on the top, and it took 23 seconds to fill the entire volume, as depicted in Fig.~\ref{fig:injection_ana}. The analysis also included the six connection holes near the bottom of the acrylic cover. A vertical weld line was observed beneath each hole as a result of the flow of molten acrylic around the hole, followed by recombination on the opposite side, as demonstrated in the right panels of Fig.~\ref{fig:injection_ana}. Since this area represents a weak point in the event of an underwater implosion of a PMT, as reported in Ref.~\cite{He:2022qzj}, a dedicated laboratory test was performed on samples strip made of thick acrylic injection molded with a 5 mm diameter hole in the center to measure the maximum tensile strength, as shown in the left panel of Fig.~\ref{fig:weld_line}. The injection gate was placed in the middle of the strip, resulting in the weld line being located on the opposite side, indicated by the dashed line. Three strips were tested using a tensile machine, as shown in the right panel of Fig.~\ref{fig:weld_line}, and the tensile force was found to be 60\% lower compared to the same strips with the central hole made by machining instead of injection molding. Based on this experiment, machining was used to produce all the holes on the acrylic covers during mass production.

\begin{figure}[ht]
\centering
  \includegraphics[width=0.7\textwidth]{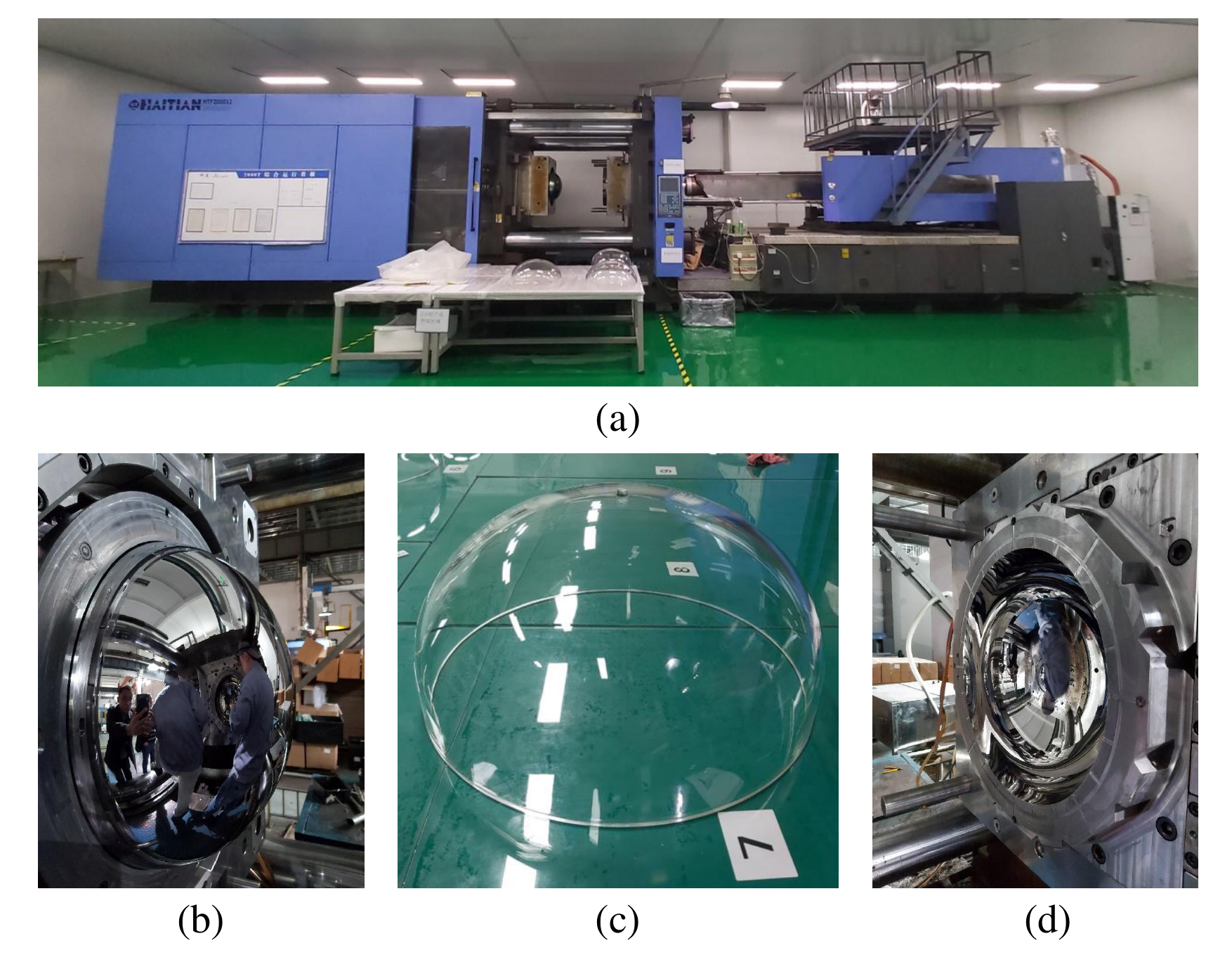}
\caption{(a) Injection machine in the clean room. (b) and (d) Injection molds. (c) One of the acrylic cover.}
\label{fig:injection}
\end{figure}

The mold comprises two parts, male and female, as depicted in Fig.~\ref{fig:injection}~(b) and (d). In order to achieve high transparency of the product, the S136 mold steel was selected for its excellent toughness, wear resistance, and corrosion resistance~\cite{s136}. The mold was meticulously machined and subsequently polished using oilstone, sandpaper, and flannel with increasingly finer abrasive paste. The roughness of the mold cannot be directly measured as it is not flat. Therefore, a sheet was created using the same mold steel and surface finish. The arithmetical mean roughness (Ra) of the sheet was measured as 0.006~$\mu$m, which meets the highest level of the standard set by the Plastics Industry Association~\cite{SPI}.

The injection molding machine was chosen with a 2,000~kton clamping force, in line with the recommendations from the mold flow analysis. A customized injection screw was manufactured to accommodate the large volume and weight of the product. Additionally, a new drying machine was acquired to precisely control the temperature and duration of drying acrylic pellets. All connection pipes were thoroughly cleaned to ensure both transparency and radiopurity. Furthermore, a class 100,000 clean room was constructed to serve as the working environment for injection, temporary storage, and pre-delivery inspection. The injection machine, along with the mold and other auxiliary equipment, is depicted in Fig.~\ref{fig:injection}~(a).

The raw acrylic pellets selected and purchased were the MGSV model from Sumitomo~\cite{Sumitomo}. Standard specimens were injection molded with this material by an inspection and certification company named SGS~\cite{sgs}, who also measured the major mechanical performance parameters presented in Table~\ref{tab:material}. The tensile modulus and tensile strength align with the specifications within 10\%, while the elongation at break is 40\% higher. Although there is no reference to Izod impact strength in the specifications, this parameter was measured as part of the standard mechanical test and was found to be comparable to casting acrylic.

\begin{table}
  \centering
  \begin{tabular}{ccccc}
    \hline
    Properties & Specification & Measurement & Casting acrylic & Standard \\
    \hline
    Tensile modulus (MPa) & 3100 & 3380 & 2400-3300 & ISO 527 (1~mm/min)\\
    Tensile strength (MPa) & 70 & 69.4 & 80 & \multirow{2}*{ISO 527 (50~mm/min)}\\
    Elongation at break (\%) & 2 & 2.8 & 2.5-4.0 & \\
    Izod impact strength (J$\cdot$m$^{-1}$) & N/A & 16 & 16-32 & ISO 179 \\
    \hline
	\end{tabular}
	\caption{Measurement of the major mechanical performance parameters for the MGSV model acrylic pellets from Sumitomo. The properties of casting acrylic are taken from Ref.~\cite{material-website}.}
	\label{tab:material}
\end{table}

Tens of acrylic covers were produced in April 2020 as a pilot production. The injection parameters were meticulously fine-tuned, primarily based on visual inspection of the product. The injection temperature was optimized between 220~$^{\circ}$C and 240~$^{\circ}$C at different sections of the injection screw, while the mold temperature was set at 70~$^{\circ}$C. The injection pressure decreased from 140~MPa to 65~MPa during the filling process and remained around 60~MPa in the mold for a few minutes after injection. The total duration of each production cycle was less than 10 minutes. The operation parameters, in particular, injection temperature and pressure, were not constant. Instead, there were adjusted within a few percent from time to time during mass production presumably due to variations of environmental temperature and humidity.

After the acrylic cover was removed from the mold, it took a few hours to cool down to room temperature, during which time the cover shrank towards the center by 0.7\%, as indicated by the mold flow analysis and the factory's experience. When the dimensions of a few products were measured on a coordinate measuring machine, the shrink rate was found to be 0.4\%, indicating that the acrylic cover was larger than the design. Specifically, the average diameter on the equatorial plane was 1.5 mm larger, making PMT installation extremely difficult. This measurement was tracked over the following two weeks and found to be stable. Considering there is a much larger clearance for the veto PMTs, we decided to produce 2,400 acrylic covers with the current mold and then modify the mold to produce the rest of the CD PMT covers with the correct dimensions. The transparency was also measured and found to be greater than 91\% in air at 413 nm. Detailed results of the dimensions and transparency will be reported in Sec.~\ref{sec.dim} and Sec.~\ref{sec.trp}.

\subsection{Drilling and chamfering}

\begin{figure}[ht]
\centering
  \includegraphics[height=0.5\textwidth]{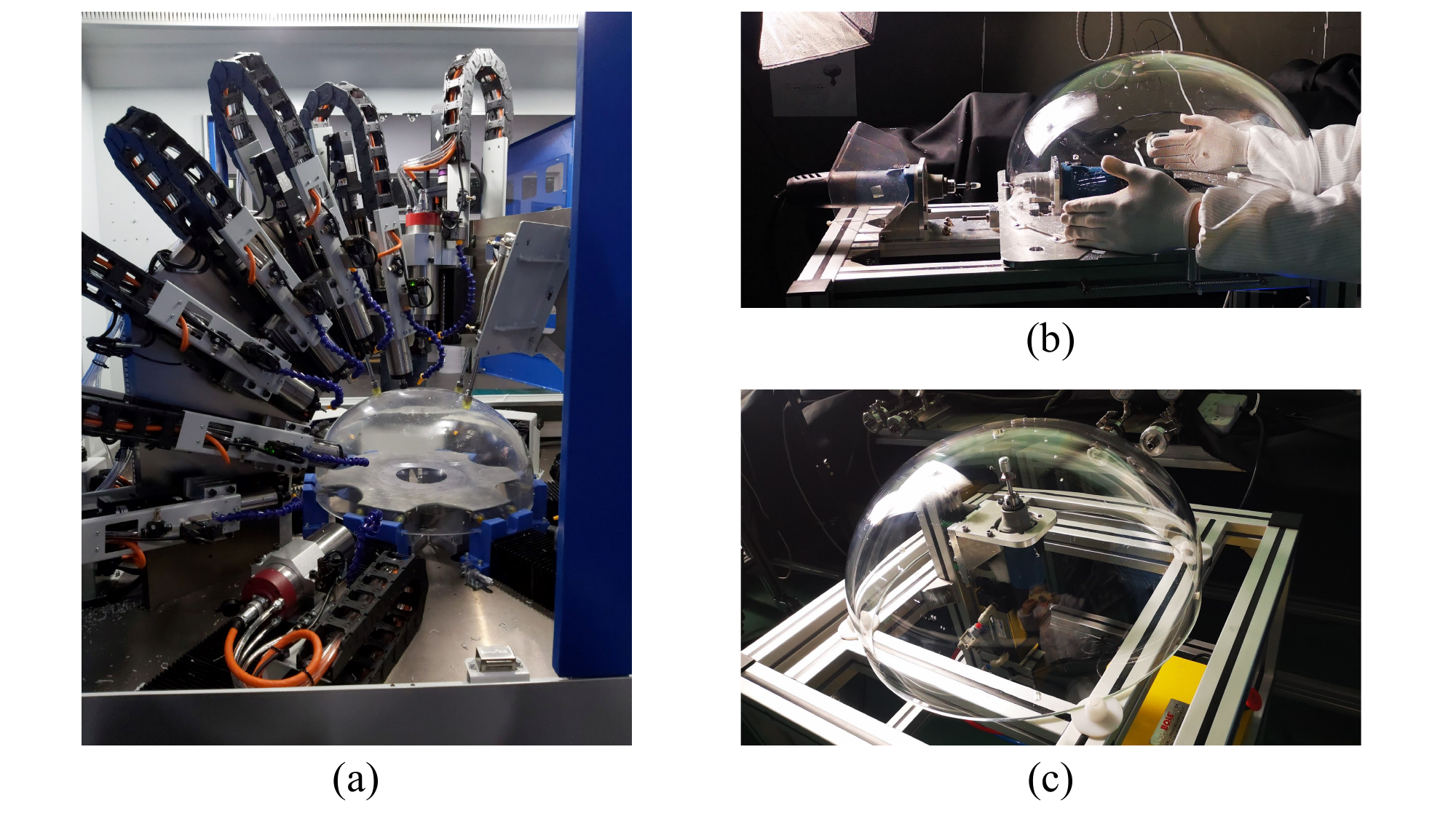}
\caption{(a) Drilling of seven inlets and six connection holes by a customized machine. (b) and (c) Chamfering of the top hole and the connection holes.}
\label{fig:drilling}
\end{figure}

A customized machine was created to drill all the necessary holes in the acrylic cover, as depicted in Fig.~\ref{fig:drilling}~(a). Seven drills were used for the vents and six drills were used for the connection holes, all made of tungsten steel. The drills were controlled by motors, with the rotating speed and feeding speed carefully adjusted to ensure a smooth inner surface of the holes. The six connection holes were drilled simultaneously, while the seven vents were drilled in two batches to avoid interference due to shorter distances. The drilling process took a few minutes.

Cracks were observed in three cases: (a) The holes were drilled only a few days after injection, leading to large stress around the top hole due to slow shrinking towards the center of the acrylic cover. (b) The water pressure in case of implosion of a PMT also created significant stress around the connection holes. (c) When the drill became dull, the drilled surface on the acrylic was unsmooth, exacerbating stress concentration. To mitigate the residual stress around these holes, several actions were taken. The drills were modified to chamfer the outer edge of the hole. Additionally, as shown in Fig.~\ref{fig:drilling}~(b) and (c), two new drills were produced to chamfer the inner edge of the top hole and all connection holes. Furthermore, all covers were required to sit for more than one month before drilling. If the surface of the hole was found to be unsmooth, the drills were promptly replaced.

After drilling and chamfering, the acrylic covers were blown with dry air to remove residual acrylic fragments, and finally were washed by pure water before being assembled to PMTs.

\section{Mass production and performance}
\label{sec.massprod}

\subsection{Mass production}

\begin{figure}[ht]
\centering
  \includegraphics[width=0.7\textwidth]{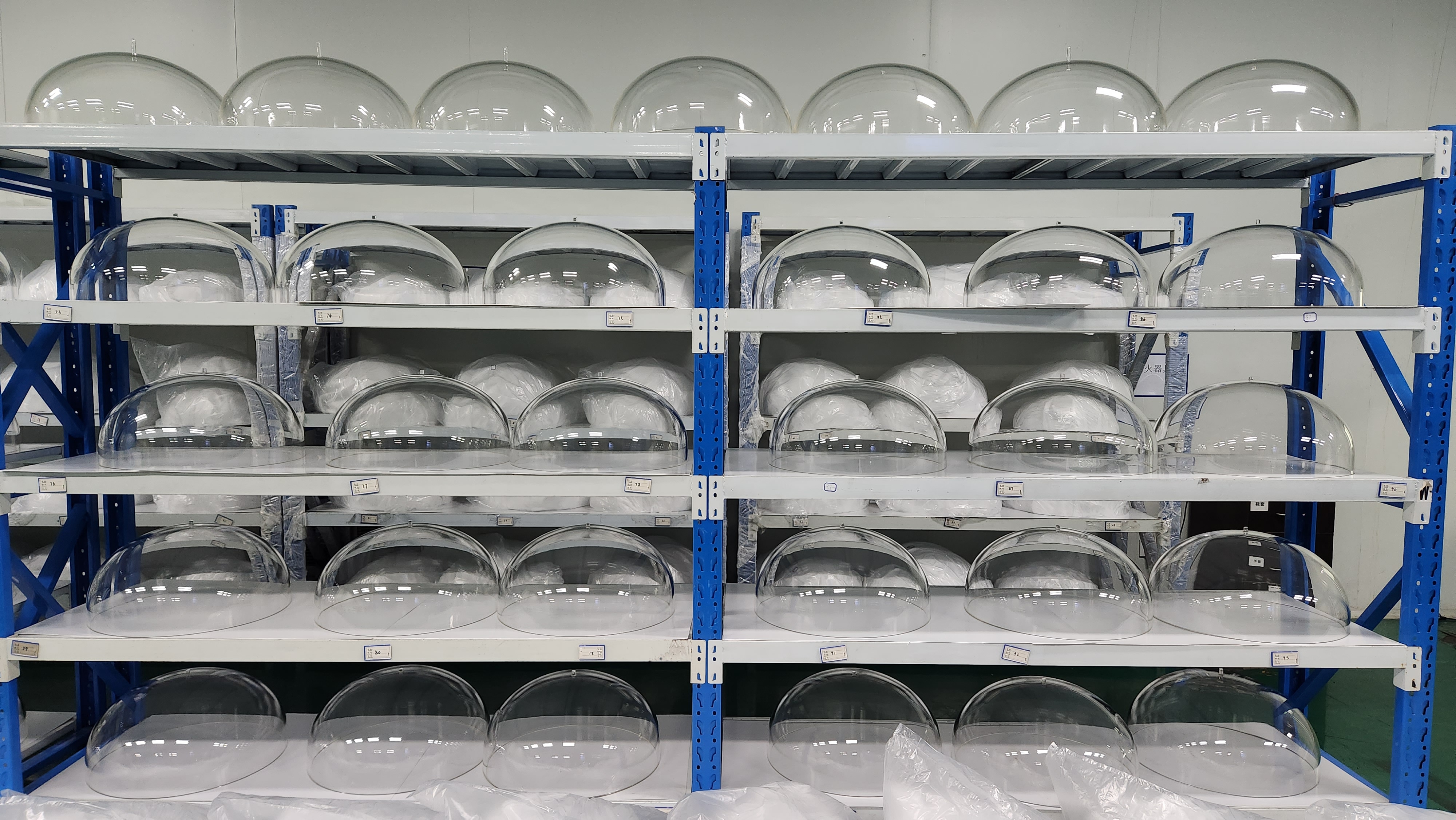}
\caption{A storage rack with injection-molded acrylic covers in the clean room.}
\label{fig:mass_prod}
\end{figure}

A standard procedure for mass production was established based on the experiences gained during the pilot production. Initially, the main body was injection molded. To maintain a stable environment and reduce non-uniformity during shrinking, the product was immediately placed on a table in front of the injection machine for two hours, then moved to a storage rack in the same clean room, as shown in Fig.~\ref{fig:mass_prod}, and left for an additional 24 hours before undergoing pre-delivery inspection, including visual inspection, measurements of transparency, dimensions, and weight. Qualified products were packed with pearl wool and transferred to a warehouse with a capacity of approximately 7,000 covers. Drilling was carried out at the earliest one month after injection. Subsequently, the covers were repacked with pearl wool and transported to the 20-inch PMTs warehouse~\cite{JUNO:2022hlz}, where they would be assembled with the PMTs and the stainless steel cover.

Injection molding was divided into five stages from 2020 to 2022, as shown in Table~\ref{tab:massp}. In the first two stages, 2,400 veto-PMT covers were produced. After the first stage, there was about a one-month gap during which we modified the injection entrance to avoid damage to the cover when it was pulled out of the mold. Following this, a pair of new molds was produced with a 1.5~mm smaller diameter for CD-PMT covers, followed by the remaining three batches of production. The initial yield was 51.0\% and was increased to 69.1\% thanks to the optimization of the injection entrance. The main reasons for rejection included bubbles, flow marks, cracks, and impurities. It is noteworthy that none of the covers were rejected due to dimensions, transparency, or weight. The measurements will be reported in the following three subsections. The yield was further improved to be larger than 90\% throughout the full period of CD-PMT covers production, mainly due to better control of injection parameters. There were two 4-month breaks mainly due to storage limitations. The total number of qualified acrylic covers is 22,916 with an average yield of 85.8\%. There was almost no damage to the acrylic cover during drilling; however, a reinspection before drilling identified another 2.2\% of unqualified products. In the end, 21,900 qualified acrylic covers were produced and delivered to JUNO.

\begin{table}
  \centering
  \begin{tabular}{ccccc}
    \hline
    Batch & Period & All products & Qualified products & Yield \\
    \hline
    1 & Jul 2020 – Aug 2020 & 2123 & 1082 & 51.0\%\\
    2 & Oct 2020 – Nov 2020 & 2008 & 1388 & 69.1\%\\
    3 & Apr 2021 – May 2021 & 5381 & 4908 & 91.2\%\\
    4 & Oct 2021 – Dec 2021 & 8282 & 7494 & 90.5\%\\
    5 & Apr 2022 – Jul 2022 & 8924 & 8044 & 90.1\%\\
    Sum & & 26718 & 22916 & 85.8\%\\
    After drilling & & & 21900 & 82.0\%\\
    \hline
	\end{tabular}
	\caption{Mass production summary.}
	\label{tab:massp}
\end{table}

\subsection{Dimensions}
\label{sec.dim}

\begin{figure}[!hbt]
  \centering
  \begin{subfigure}{.45\textwidth}
  \centering
  \includegraphics[width=1\textwidth]{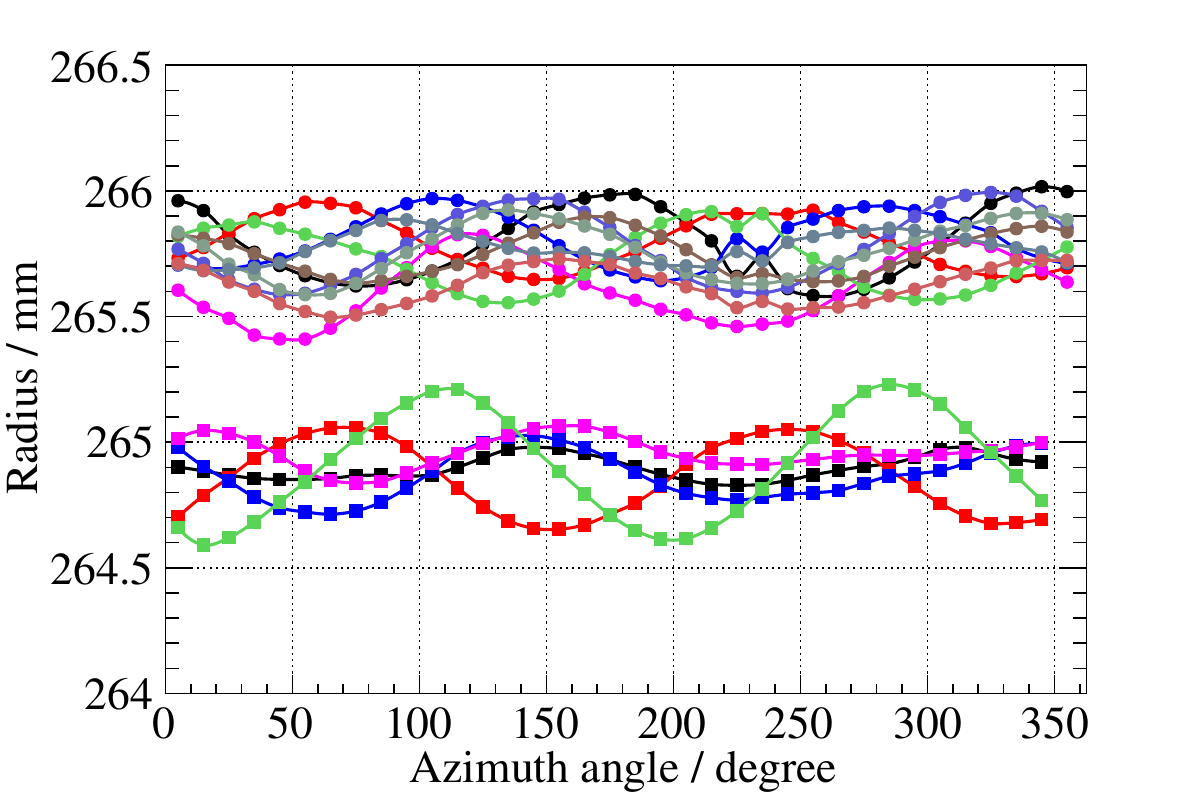}
  \caption{}
\end{subfigure}
\begin{subfigure}{.45\textwidth}
  \centering
  \includegraphics[width=1\textwidth]{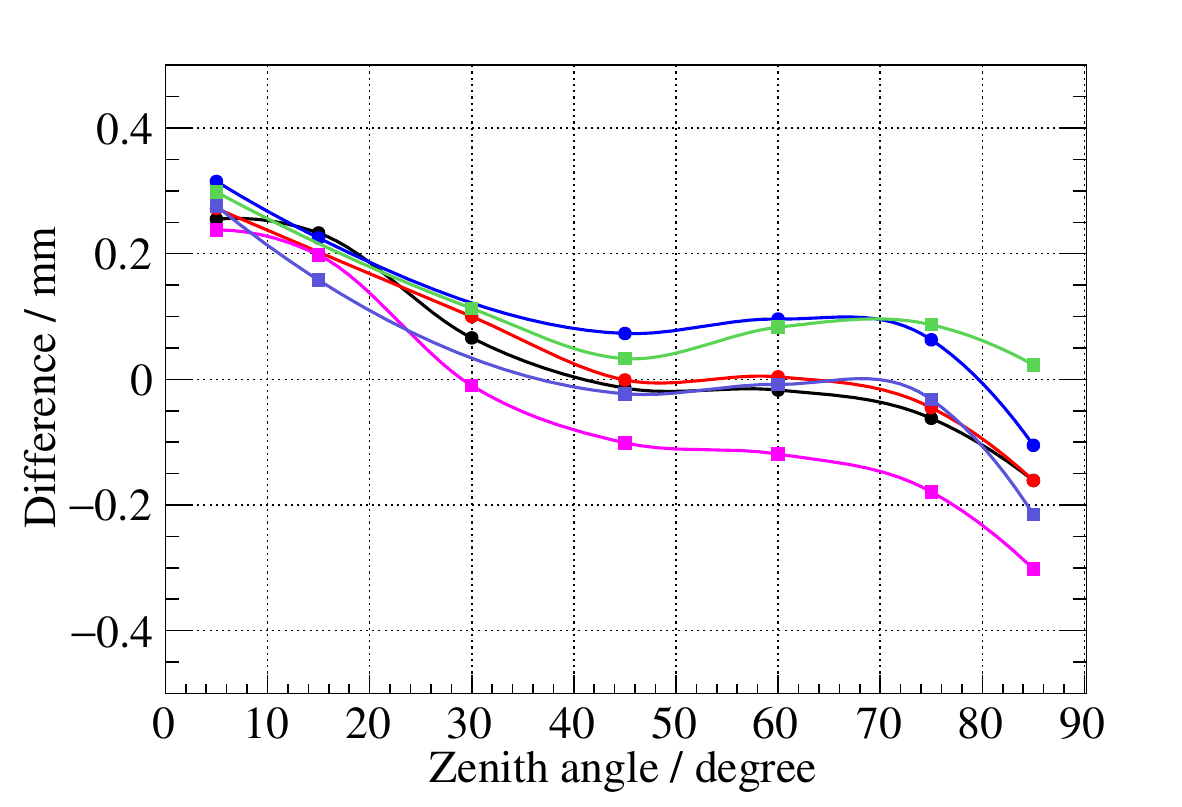}
  \caption{}
\end{subfigure}
  \caption{3D profile scanning on a coordinate measuring machine. (a) Outer radius measurement surround the bottom of the acrylic cover with an interval of 10$^{\circ}$. Markers and colors represent different acrylic covers. The ten covers with larger radius will be used for veto PMTs and the other five for CD PMTs. (b) Difference between measured and designed inner radii with 5$^{\circ}$ or 10$^{\circ}$ interval along different meridians. The circular and square markers represent two acrylic covers.}
  \label{fig:3d}
\end{figure}

In the early stage of mass production, ten veto-PMT covers and five CD-PMT covers were randomly selected for measurement on a coordinate measuring machine. At 10$^{\circ}$ intervals along the equatorial plane of the cover, 35 or 36 points were measured, with each point's distance to the center representing the measured radius. The results are shown in Fig.~\ref{fig:3d}~(a). Each cover exhibited a clear oscillation structure, indicating an ellipse with the two peaks corresponding to the long axis and the two valleys corresponding to the short axis. The average radius on the equatorial plane of the veto-PMT covers was 265.7~mm with a spread of $\pm$0.4~mm, and of the CD-PMT covers was 264.9~mm with a spread of $\pm$0.4~mm. In the later production stage, profiles of two CD-PMT covers were further studied by measuring the inner radius on the semi-ellipsoid along three meridians separated by 120$^{\circ}$ and comparing them with the design. The results are shown in Fig.~\ref{fig:3d}~(b). The maximum difference was less than 0.4~mm due to shrinking.

The thickness of the acrylic cover was examined by randomly selecting 24 veto-PMT covers and 43 CD-PMT covers. Measurements were taken using a 0.1-mm precision caliper at the top through one of the water inlets and at the bottom, and the results are summarized in Table~\ref{tab:thickness}. It was observed that the thickness of the veto-PMT covers was 0.1-0.3~mm larger compared to the CD-PMT covers. However, both types of covers were found to be consistent with the design, with measurements falling within $\pm$0.23~mm of the specified thickness and a maximum standard deviation of 0.15~mm.

Based on these measurements, a dedicated tool was created as a circular groove with an inner radius of 255.5~mm and an outer radius of 265.5~mm to verify the bottom dimension of every CD-PMT cover during mass production. Therefore, the width of the groove is 10~mm. Considering the product thickness is 9~mm on the bottom, the total tolerance for both the radius and the thickness needs to be less than 1~mm. For the veto-PMT cover, where the installation clearance is much larger, the verification tool was also utilized, but with a slightly relaxed tolerance range of 255.9~mm and 266.1~mm for the inner and outer radius, respectively.

\begin{table}
  \centering
  \begin{tabular}{cccccc}
    \hline
    \multirow{2}*{Type} & \multirow{2}*{Quantities} &
    \multicolumn{2}{c}{Thickness on the bottom (mm) } & \multicolumn{2}{c}{Thickness on the top (mm)} \\
    & & Average & Standard deviation & Average & Standard deviation \\
    \hline
    Veto-PMT cover & 24 & 9.23 & 0.05 & 11.01 & 0.08 \\
    CD-PMT cover & 43 & 8.98 & 0.10 & 10.91 & 0.15 \\
    \hline
	\end{tabular}
	\caption{Measurements of the acrylic cover thickness on the bottom and on the top with randomly selected products.}
	\label{tab:thickness}
\end{table}

\subsection{Transparency}
\label{sec.trp}

\begin{figure}[!hbt]
  \centering
  \begin{subfigure}{.45\textwidth}
  \centering
  \includegraphics[width=0.9\textwidth]{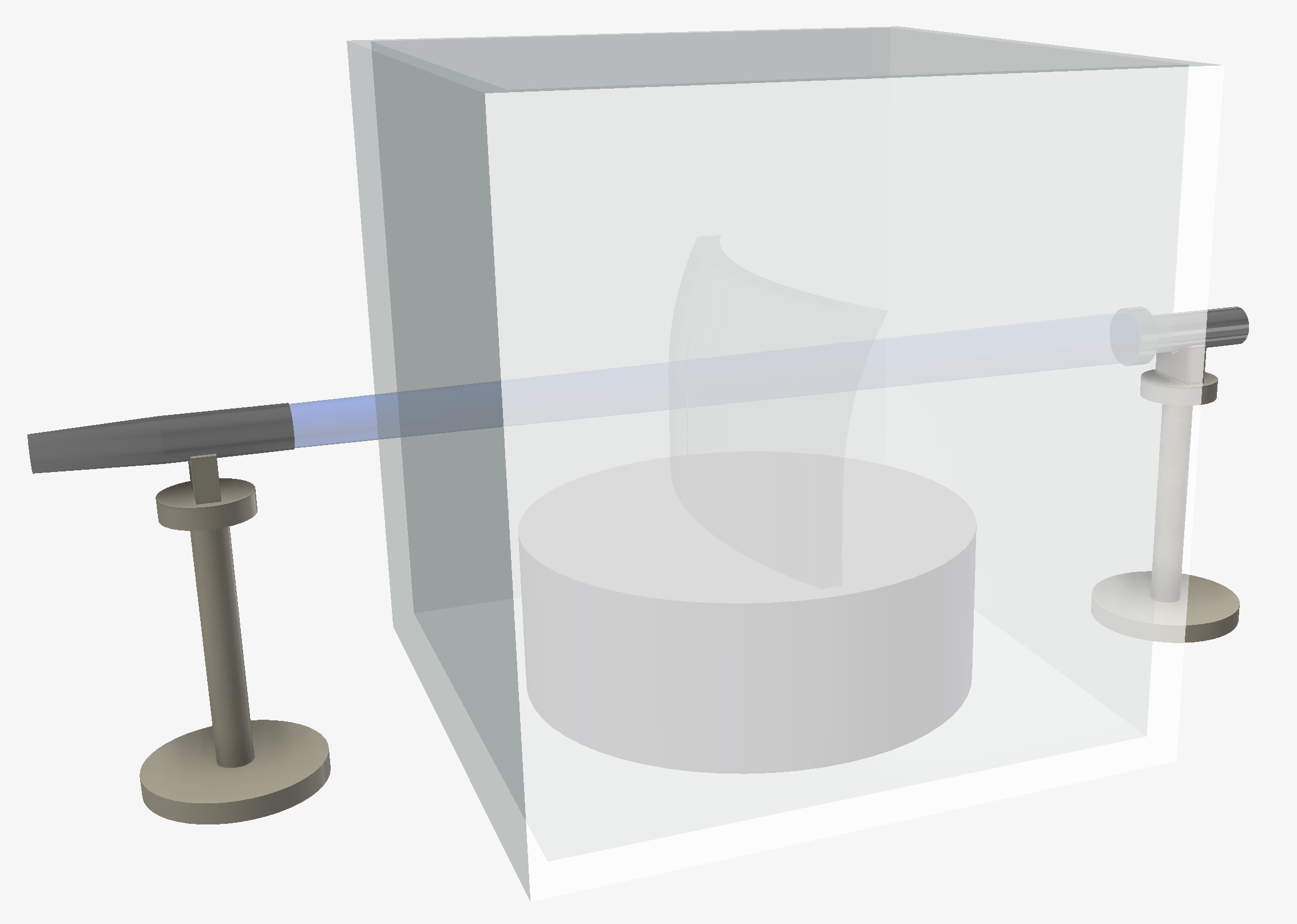}
  \caption{}
\end{subfigure}
\begin{subfigure}{.45\textwidth}
  \centering
  \includegraphics[width=1\textwidth]{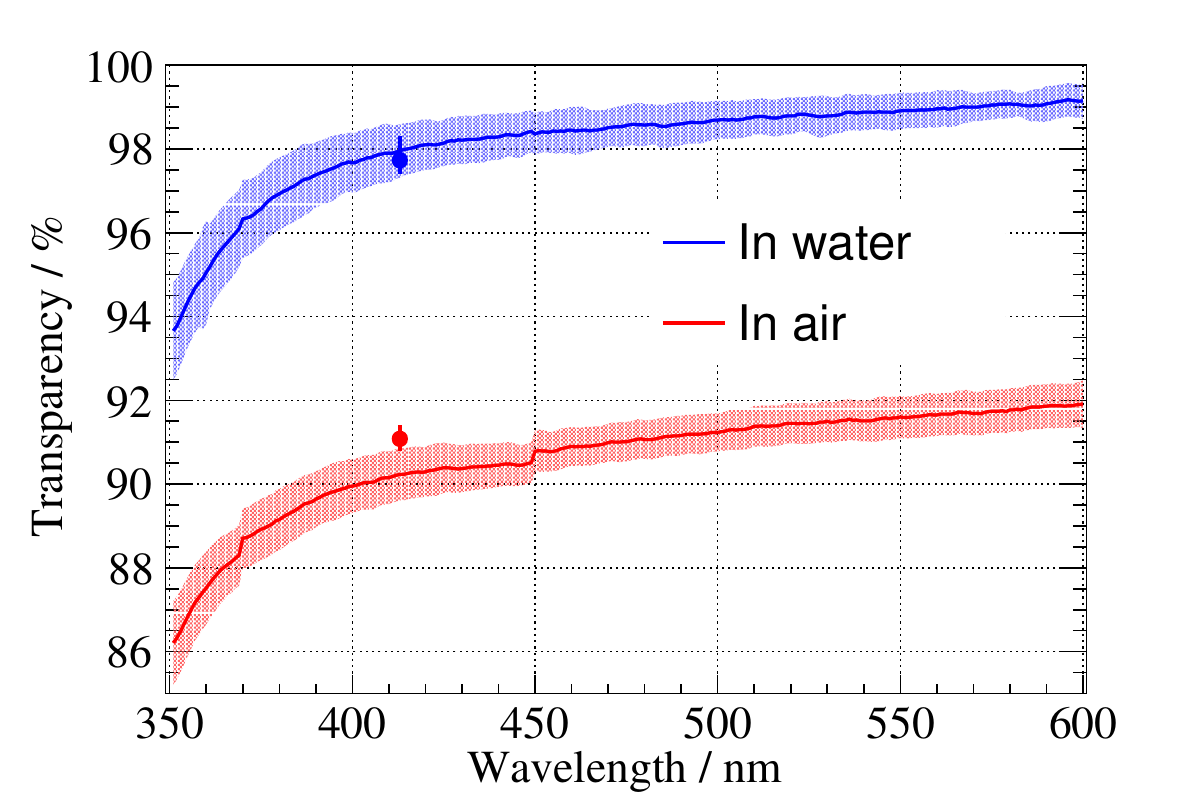}
  \caption{}
\end{subfigure}
  \caption{(a) Schematic diagram of the setup for transparency measurement. (b) Transparency as a function of wavelength by the spectrometer. The two bands encapsulate results of twelve samples and the two lines are their averages. The blue and red data points, along with error bars, represent the measurements obtained from the laser system at 413~nm.}
  \label{fig:transparency}
\end{figure}

The commercial spectrometer JH723PC~\cite{Spectrometer} with a precision of 0.3\% was utilized to measure the transparency of the acrylic cover in water within the most interesting spectral range of JUNO between 350~nm and 600~nm, with a scanning interval of 1~nm. The light spot in the spectrometer is about 20~mm$\times$5~mm. Due to space limitations within the spectrometer, two acrylic covers were cut into twelve 5~cm$\times$6~cm panels from different positions. Each panel was affixed to a flat base, placed in a transparent vessel filled with pure water, and positioned in the spectrometer. The panel was adjusted to be perpendicular to the light beam, as illustrated in Fig.~\ref{fig:transparency}~(a). Transparency at a specific wavelength was determined by comparing the measured relative light intensity with and without the panel. This process was repeated for all twelve panels, and the results are depicted as the blue line and band in Fig.~\ref{fig:transparency}~(b). The average transparency at 420~nm was found to be 98.1\%, with an uncertainty of 0.5\%, estimated by considering the precision of the spectrometer, panel adjustment, and variations among different panels. Transparency in air was also measured and was approximately 8\% lower at 420~nm compared to the result in water, primarily due to greater reflection at the two boundaries of acrylic and air. Small fluctuations at 370~nm and 450~nm correspond to the switching of the optical grating in the spectrometer.

Due to the destructive nature of using a spectrometer for measuring the acrylic cover, a customized laser system~\cite{Yang:2020juno} was employed. This system provides a single-wavelength laser at 413~nm, with a spot diameter of approximately 5~mm and a power of around 100~mW, along with a laser power meter. The schematic diagram is similar to that in Fig.~\ref{fig:transparency}~(a). In contrast to the approach described in Ref.~\cite{Yang:2020juno}, where the sample was fixed and the laser system was in motion, in this measurement, the laser generator and the power meter were aligned and fixed on an optical table at a distance of approximately 50~cm. The sample being measured was positioned 1-2~cm away from the power meter to minimize the impact of light scattering. The laser powers obtained with and without the acrylic were compared to determine the transparency. This measurement typically took less than one minute in natural light. The system's stability was monitored by comparing the laser power before and after measuring the acrylic cover, and it was found to be less than 0.2~mW.

The acrylic panels mentioned above were measured in both water and air using the laser system, depicted as the blue and red dots in Fig.~\ref{fig:transparency}~(b). The results at 413~nm in water are consistent, whereas the transparency measured by the spectrometer in air is 0.9\% lower than that obtained by the laser system. This difference is likely due to the larger light spot, resulting in greater divergence of light reflection in the ellipsoid panel.

\begin{figure}[ht]
\centering
    \includegraphics[width=0.5\textwidth]{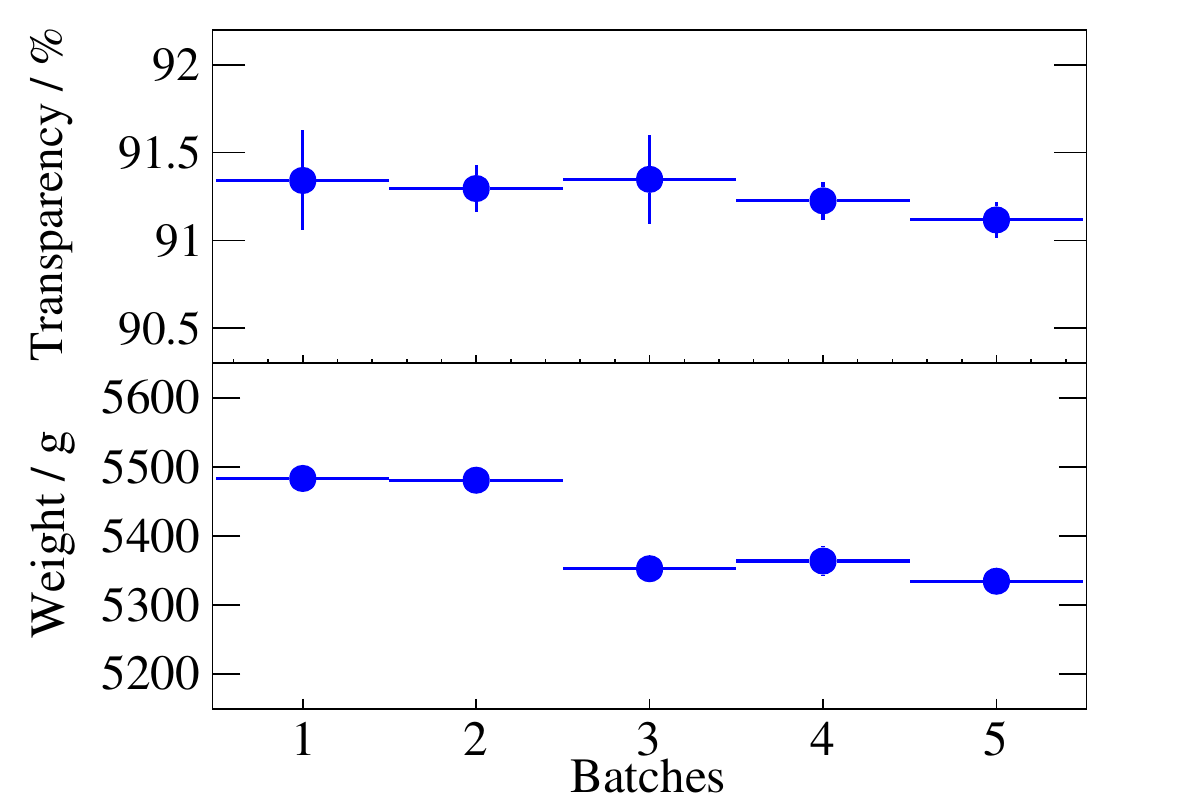}
\caption{Average transparency at 413~nm in air (top panel) and weight (bottom panel) for different production batches defined in Table~\ref{tab:massp}. The error bars represent the standard deviation.}
\label{fig:batch}
\end{figure}

During mass production, the laser system was utilized to monitor the transparency of the acrylic covers through a 10\% sampling measurement in air. The average transparency of all 2,411 measured acrylic covers was 91.3\%, with a standard deviation of 0.2\%, consistent with the panel measurement of 91.1\% in Fig.~\ref{fig:transparency}~(b). This information is further presented for each production batch in the top panel of Fig.~\ref{fig:batch}. The results remained stable throughout the entire production period, with a maximum difference of 0.2\%.

\subsection{Weight}

The weight of all the acrylic covers was measured for quality control and as a reference for mechanical performance, as depicted in the bottom panel of Fig.~\ref{fig:batch}. The average weight of the first two batches of veto-PMT covers is 5493~g, with a standard deviation of 21~g, while the average weight of the last three batches of CD-PMT covers is 5350~g, with a standard deviation of 22~g. This 2.7\% difference aligns with the 1\%-3\% variation in thickness between these two types of acrylic covers. Additionally, the 0.3\% larger diameter of veto-PMT covers also contributes to this distinction.

\subsection{Underwater experiments}

\begin{figure}[ht]
\centering
\begin{subfigure}{.35\textwidth}
  \centering
  \includegraphics[width=.95\textwidth]{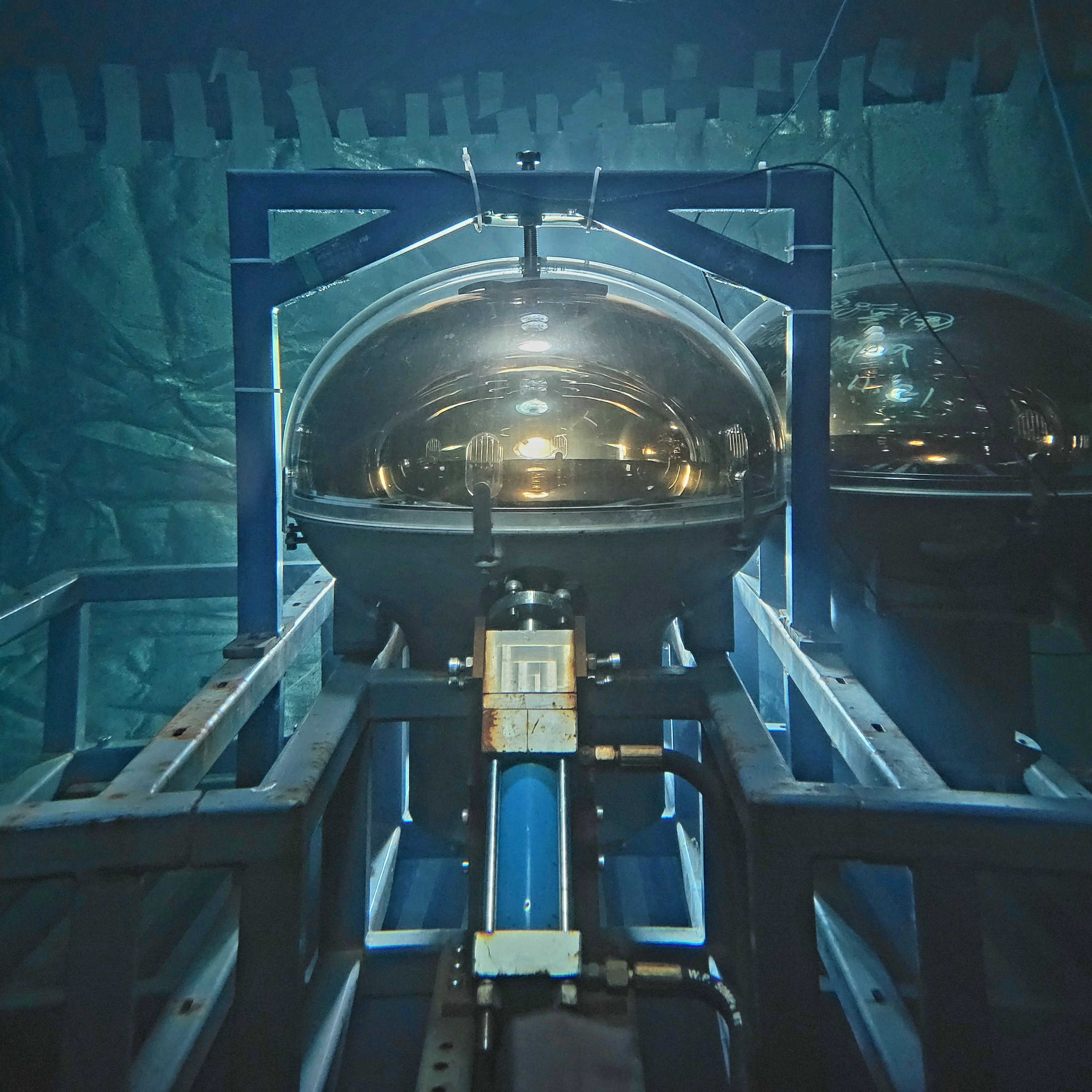}
  \caption{}
\end{subfigure}
\begin{subfigure}{.35\textwidth}
  \centering
  \includegraphics[width=.95\textwidth]{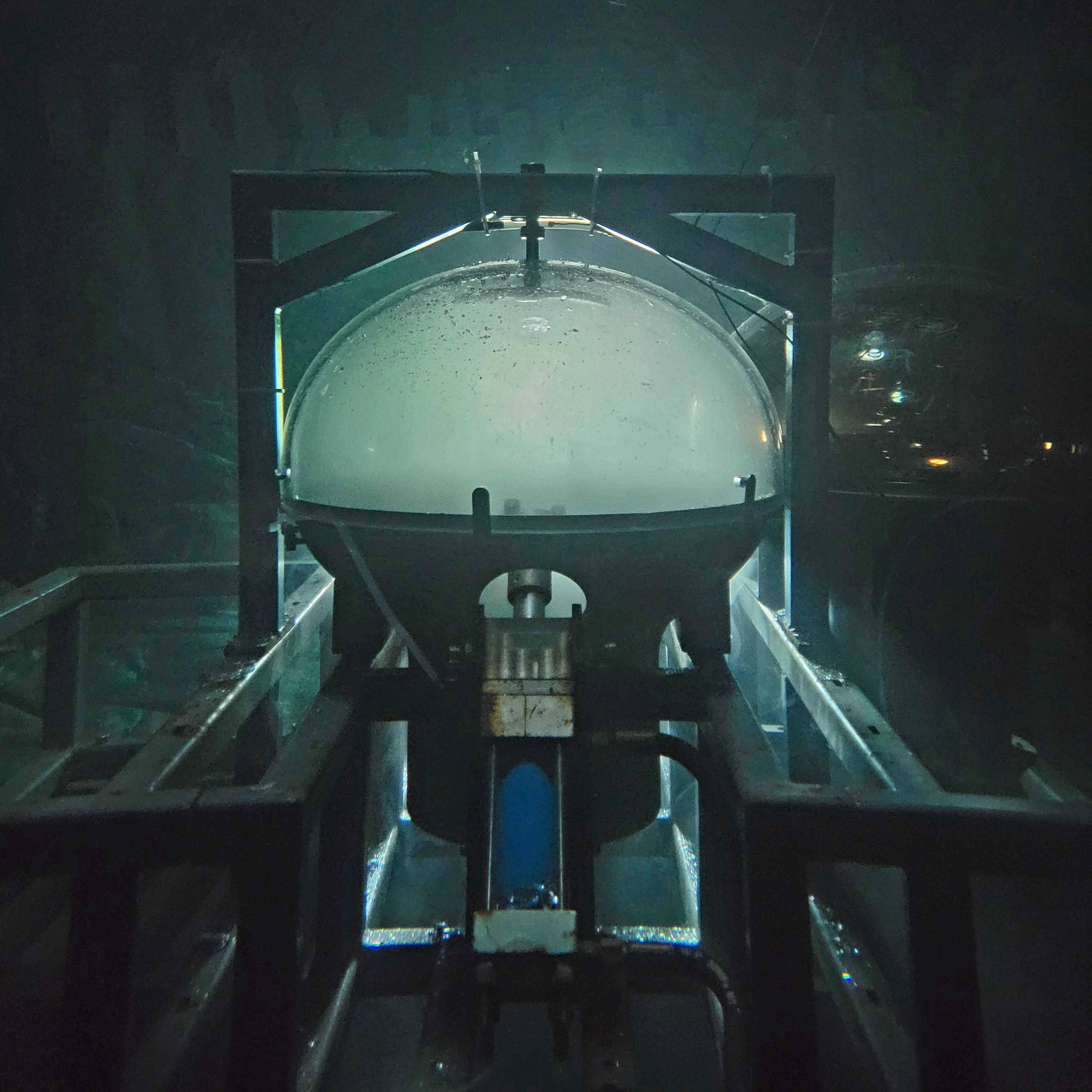}
  \caption{}
\end{subfigure}
\caption{An underwater experiment (a) before and (b) after imploding the central PMT.}
\label{fig:test25}
\end{figure}

During mass production, three underwater experiments were performed to validate the protection structure. The experimental setup was consistent with that in Ref.~\cite{He:2022qzj}, with both the acrylic cover and the stainless steel cover randomly selected from different production stages each time. Specifically, one acrylic cover was used for the veto PMT, and the other two were for the CD PMTs. In Fig.~\ref{fig:test25}, only the third experiment is presented as an example. Two PMTs were assembled with the protection structure in a high-pressure tank filled with water, with a few cubic meters of air remaining in the tank. The pressure was increased to 0.5~MPa above atmospheric pressure using an air pump. The picture of the experimental setup is shown in Fig.~\ref{fig:test25}~(a). Initiating the implosion, the central PMT (the left one in golden in the picture) was shattered into pieces by three stainless steel plates, leading to its disappearance in Fig.~\ref{fig:test25}~(b). Despite a few small cracks around some of the connection holes caused by the high water pressure, the protection cover remained intact. There was no damage to the neighboring PMT (the right one in Fig.~\ref{fig:test25}~(b)) or its protection cover, thereby preventing a chain reaction of implosion.

The other two experiments yielded similar results. As detailed in Ref.~\cite{He:2022qzj}, the presence of the protection cover significantly slowed down the water inflow and reduced the shockwave by a factor of 50. This demonstrates the robustness of the mass-produced acrylic covers.

\section{Summary}
\label{sec.summary}

A protection system was developed to prevent underwater implosion chain reactions of JUNO PMTs. As a key component of the protection system, the acrylic cover was designed as a hollow structure approximately 500~mm in size, with a thickness of around 10~mm and a weight of about 5.4 kg. Over a period of more than two years, more than 20,000 acrylic covers were produced.

The main structure of the acrylic covers was created using injection molding technology, with all the necessary holes drilled to avoid weld lines. A 2,000 kton clamping force injection molding machine was utilized in a class 100,000 clean room to minimize dust contamination. The mold was machined and highly polished up to a roughness of 0.006~$\mu$m to improve the final acrylic transparency. The injection process took less than 10 minutes, followed by additional time for drilling and cleaning. Injection parameters were optimized during pilot production, revealing a shrink ratio of the acrylic that was 0.3\% smaller than expected, resulting in a diameter 1.5 mm larger than anticipated. Consequently, 2,400 acrylic covers were initially produced and will be used in the water Cherenkov detector, which has a much larger clearance between two PMTs than in the Central Detector. A new mold was subsequently created for the remaining acrylic covers, ensuring the correct dimensions.

The initial yield in mass production was 51\%, but it increased and stabilized beyond 90\% due to the optimization of the injection inlet and improved control of injection parameters. The average yield reached 82\%. Profiles of randomly sampled acrylic covers were analyzed using a coordinate measuring machine, revealing a precision of 0.2\% in terms of the radius at different positions. Thicknesses at the top and the bottom were also measured, with the standard deviation found to be smaller than 0.2~mm. The transparency of two acrylic cover fragments was measured in water using a spectrometer between 350~nm and 600~nm, resulting in a transparency of 98.1\%$\pm$0.5\% at 420~nm, consistent with the results obtained by a laser system with a single wavelength. Throughout the full production period, the dimensions, transparency, and weights of the acrylic covers were monitored by customized tools and found to be stable. Finally, the mechanical performance was validated through three 0.5~MPa underwater experiments.

\section{Acknowledgments}
This work was supported by the Strategic Priority Research Program of the Chinese Academy of Sciences, Grant No. XDA10011100, and People's Government of Haiyan County.
We thank Dr. Xiaoyu Yang for advising the laser system for the transparency monitoring during acrylic covers mass production. We thank North Night Vision Technology Co., Ltd for providing all PMTs in the underwater experiments. We thank Naval Research Academy for providing the high-pressure water tank and for helping the underwater experiments.

\bibliographystyle{h-physrev5}
\bibliography{references}

\begin{thebibliography}{10}

\bibitem{Super-Kamiokande:2002weg}
Super-Kamiokande, Y.~Fukuda {\em et~al.}, {The Super-Kamiokande detector},
\newblock Nucl. Instrum. Meth. A {\bf 501}, 418 (2003).

\bibitem{Borexino:2008gab}
Borexino, G.~Alimonti {\em et~al.}, {The Borexino detector at the Laboratori
  Nazionali del Gran Sasso},
\newblock Nucl. Instrum. Meth. A {\bf 600}, 568 (2009), arXiv:0806.2400.

\bibitem{KamLAND:2002uet}
KamLAND, K.~Eguchi {\em et~al.}, {First results from KamLAND: Evidence for
  reactor anti-neutrino disappearance},
\newblock Phys. Rev. Lett. {\bf 90}, 021802 (2003), arXiv:hep-ex/0212021.

\bibitem{Super-Kamiokande:2010tar}
Super-Kamiokande, K.~Abe {\em et~al.}, {Solar neutrino results in
  Super-Kamiokande-III},
\newblock Phys. Rev. D {\bf 83}, 052010 (2011), arXiv:1010.0118.

\bibitem{An:2015jdp}
JUNO, F.~An {\em et~al.}, {Neutrino Physics with JUNO},
\newblock J. Phys. G {\bf 43}, 030401 (2016), arXiv:1507.05613.

\bibitem{Djurcic:2015vqa}
JUNO, Z.~Djurcic {\em et~al.}, {JUNO Conceptual Design Report},
\newblock (2015), arXiv:1508.07166.

\bibitem{JUNO:2022hxd}
JUNO, A.~Abusleme {\em et~al.}, {JUNO physics and detector},
\newblock Prog. Part. Nucl. Phys. {\bf 123}, 103927 (2022).

\bibitem{He:2022qzj}
M.~He {\em et~al.}, {Design of the PMT underwater cascade implosion protection
  system for JUNO},
\newblock JINST {\bf 18}, P02013 (2023), arXiv:2209.08441.

\bibitem{s136}
\url{https://www.otaisteel.com/s136-mold-steel/}.

\bibitem{SPI}
\url{https://www.spifinish.com/technology}.

\bibitem{Sumitomo}
\url{http://www.sumitomo-chem.com.sg}.

\bibitem{sgs}
\url{https://www.sgs.com/}.

\bibitem{material-website}
\url{http://www.goodfellow.com}.

\bibitem{JUNO:2022hlz}
JUNO, A.~Abusleme {\em et~al.}, {Mass Testing and Characterization of 20-inch
  PMTs for JUNO},
\newblock (2022), arXiv:2205.08629.

\bibitem{Spectrometer}
\url{http://www.jh17.cn/english}.

\bibitem{Yang:2020juno}
X.~Yang {\em et~al.}, {The measurement system of acrylic transmittance for the
  JUNO central detector},
\newblock Rad. Det. Tech. Meth. {\bf 4}, 284 (2020).

\end{thebibliography}

\end{document}